\documentclass{article}

\usepackage{arxiv}

\usepackage[utf8]{inputenc} 
\usepackage[T1]{fontenc}    
\usepackage[colorlinks=true, linkcolor=blue, citecolor=blue, urlcolor=blue, pdfborder={0 0 0}]{hyperref}
\usepackage{url}            
\usepackage{booktabs}       
\usepackage{amsfonts}       
\usepackage{nicefrac}       
\usepackage{microtype}      
\usepackage{lipsum}		
\usepackage{graphicx}
\usepackage{doi}
\usepackage{float}
\usepackage{amsthm}
\usepackage{physics}
\usepackage{bm}

\usepackage{longtable}
\usepackage{array}     
\usepackage{lmodern}   

\theoremstyle{definition}

\newcommand{\supplementaryfigures}{
  \renewcommand{\thefigure}{S\arabic{figure}}
  \setcounter{figure}{0}
}
\newcommand{\supplementarytables}{
  \renewcommand{\thetable}{S\arabic{table}}
  \setcounter{table}{0}
}

\newcommand{\one}{I}
\newcommand{\two}{\mbox{I\hspace{-1.2pt}I}}
\newcommand{\twoo}{I\hspace{-0.5pt}I}





\title{Two universal pathways in demographic transition}

\author{Kenji Itao\\
	Computational Group Dynamics Collaboration Unit,\\
    RIKEN Center for Brain Science, \\
    2-1 Hirosawa, Wako, Saitama 351-0198, Japan.\\
	\texttt{kenji.itao@riken.jp}
}

\renewcommand{\shorttitle}{Two universal pathways in demographic transition}

\hypersetup{
pdftitle={Two universal pathways in demographic transition},
pdfauthor={Kenji Itao},
pdfkeywords={demographic transition, global data analysis, universality, statistical physics},
}

\begin{document}
\maketitle

\begin{abstract}
Demographic transition, characterized by declines in fertility and mortality, is a global phenomenon associated with modernization. While typical patterns of fertility decline are observed in Western countries, their applicability to other regions and the underlying mechanisms remain unclear. By analyzing demographic data from 195 countries over 200 years, this study identifies two universal pathways in the changes in the crude birth rate (i.e., births per 1,000 individuals $\lambda$) and life expectancy at birth ($e_0$), characterized by the conservation of either $\lambda e_0$ or $\lambda \exp(e_0 / 18)$. These pathways define two distinct phases governed by different mechanisms. Phase \one, characterized by the conservation of $\lambda e_0$, dominated until the mid-20th century, with high child mortality and steady population growth. In contrast, Phase \twoo, conserving $\lambda \exp(e_0 / 18)$, has prevailed since 1950, featuring low child mortality and steady GDP per capita growth. A theoretical model considering the trade-off between reproduction and education elucidates the transition between these phases, demonstrating that population size is prioritized in Phase \one, while productivity is maximized in Phase \twoo. Modernization processes, such as declining educational costs and increasing social mobility, are identified as key accelerators of the transition to Phase \twoo. The findings suggest that reducing educational costs can foster fertility recovery without compromising educational standards, offering potential policy interventions. This study provides a novel theoretical framework for understanding demographic transition by applying principles from statistical physics to uncover universal macroscopic laws and their underlying mechanisms.
\end{abstract}

\keywords{social institutions \and evolutionary game theory, \and dynamical-systems game \and statistical physics}

\section*{Introduction}
Declines in fertility and mortality, commonly referred to as the demographic transition, are widely observed in modern societies \cite{notestein1945population, kirk1996demographic}. Modernization typically initiates this transition with a decline in mortality, leading to rapid population growth, followed by a later decrease in fertility and, more recently, population shrinkage \cite{kirk1996demographic, bongaarts2009human, reher2012population}. The economic, social, and political implications of this transition have been examined in demography, economics, and history, while also drawing considerable attention from policymakers and the public alike \cite{bloom2003demographic, reher2012population, murphy2017economization}.

Researchers have long explored the typical patterns of demographic transition. Demographers have identified two stages of demographic transition in Western countries. The initial decline in fertility to the replacement level (i.e., two children per woman) is termed the first demographic transition, whereas subsequent declines below the replacement level are referred to as the second demographic transition \cite{lesthaeghe2010unfolding, lesthaeghe2014second, zaidi2017second}. 
The shift to the second demographic transition is associated with lower marriage rates, delayed childbearing, the pursuit of ``higher-order needs'' such as education and self-realization, and increased female economic empowerment, among other factors \cite{kirk1996demographic, mulder1998demographic, mace2000evolutionary, bongaarts2009human, lesthaeghe2014second, zaidi2017second, lutz2019education}. 
Similarly, the ``unified growth theory'' proposes that modern technological advancements drive the transition from the Malthusian growth phase—characterized by stable population growth—to the modern growth phase, marked by sustained GDP per capita growth \cite{galor2011unified}. These frameworks suggest the potential universality of demographic dynamics across countries. However, the extent to which such universality applies globally remains an open question, necessitating a comprehensive quantitative investigation \cite{lesthaeghe2014second}.

Additionally, the mechanisms driving demographic transitions are controversial; although various disciplines have proposed explanations, the fundamental drivers remain elusive \cite{kirk1996demographic, lesthaeghe2010unfolding, lesthaeghe2014second, zaidi2017second, galor2011unified, ihara2004cultural, cervellati2005human, sear2015evolutionary, colleran2016cultural}. The decline in mortality is generally attributed to improvements in nutrition, infrastructure, and medicine \cite{kirk1996demographic, canning2011causes}. In contrast, explanations for declining fertility span multiple perspectives. Demographers highlight population pressure \cite{malthus1798essay}, cultural shifts toward postmodern norms \cite{lesthaeghe2010unfolding, lesthaeghe2014second, zaidi2017second, ihara2004cultural}, the implementation of family planning programs \cite{amin2002spatial, murphy2017economization}, and human capital accumulation through education \cite{galor2011unified, canning2011causes}.
Economists emphasize how individuals optimize fertility decisions by balancing reproduction and educational investment—commonly referred to as the quantity-quality trade-off \cite{hanushek1992trade, bleakley2009chronic, becker2010trade, werding2014children, fernihough2017human}—and by weighing career advancement and leisure preferences against childbearing \cite{hakim2003new, vitali2009preference, galindev2011leisure}. Human behavioral ecologists, meanwhile, focus on maximizing children’s reproductive value through investments in embodied capital, primarily via education \cite{kaplan1996theory, mace2000evolutionary, lawson2011parental, shenk2009testing, sear2015evolutionary, colleran2016cultural}. Comparative studies indicate that childbearing in urban societies is costlier, whereas the returns to educational investment are higher than traditional settings \cite{kaplan1996theory, mace2008reproducing, colleran2015fertility}.
Finally, cultural evolutionists propose that although the preference for smaller family sizes may be maladaptive, it is widely transmitted due to social learning biases and group-level cultures \cite{mulder1998demographic, ihara2004cultural, richerson2008not, newson2007influences, sear2015evolutionary, colleran2016cultural}.

To uncover universal patterns and clarify the fundamental mechanisms of demographic transition, it is essential to identify macroscopic measures of demographic dynamics. Different driving forces produce distinct macroscopic outcomes. For example, when fertility decline is primarily driven by population pressure, the total population size (or its growth rate) remains conserved. Population growth can be quantified by the ratio of the crude birth rate, $\lambda$, to life expectancy at birth, $e_0$, under stationary conditions \cite{preston2000demography}. Accordingly, in a population-pressure-driven scenario, the product $\lambda e_0$ remains constant. In contrast, if other factors such as educational investment drive fertility decline, $\lambda e_0$ will no longer be conserved, and different quantities may remain constant instead. Identifying such conserved quantities is crucial for elucidating the mechanisms underlying demographic transitions and assessing their universality.

By analyzing global data, this study demonstrates that demographic transition comprises two distinct phases, each characterized by different conserved quantities and underlying mechanisms. The universality of these phases suggests that demographic transitions across global countries can be explained by two fundamental mechanisms, regardless of political systems or historical contexts. 
A simple model, considering the trade-off in parental investment between fertility and education, elucidates these mechanisms, which maximize either total population size or total productivity. By quantitatively characterizing empirical trends and their mechanisms, this study aims to establish an integrative framework for understanding demographic transition to place existing explanations in a broader perspective.

In the following sections, I examine the relationship between the crude birth rate\footnote{In this study, I use the crude birth rate as a measure of fertility rather than the total fertility rate (TFR), which is commonly employed in demographic research \cite{bongaarts2009human, lesthaeghe2014second}. The TFR is calculated as the sum of age-specific fertility rates for women aged 15--49 \cite{preston2000demography}, showing the hypothetical number of children a woman would have if she survived beyond age 50. Notably, a TFR of 2 indicates a stable population size only if all women survive past 50. In contrast, the crude birth rate, which directly counts the number of births, provides a more immediate measure of population dynamics. Nonetheless, Fig.S1 demonstrates that the qualitative trends are similar when using TFR.}(\(\lambda\)), and life expectancy at birth (\(e_0\)) across 195 countries from 1800 to 2015, using Gapminder data\footnote{For consistency, I also analyzed data from the Human Mortality Database \cite{HMD} and the United Nations Statistics Division \cite{UNSD}, as shown in Fig.S2.}. This analysis reveals two universal pathways of demographic transition, marked by different conserved quantities, on which individual countries' trajectories of \(\lambda\) and \(e_0\) converge. Specifically, I identify two universal phases: Phase \one, conserving $\lambda e_0$, and Phase \two, conserving $\lambda \exp(e_0 / 18)$. 
Each phase is characterized by additional demographic metrics. This analysis links these phases to established demographic patterns, while partially updating existing frameworks, particularly on demographic transitions in developing countries \cite{lesthaeghe2010unfolding, lesthaeghe2014second, zaidi2017second, galor2011unified}.
Finally, I propose a simple model to elucidate the mechanisms driving these two phases and to explain why these quantities are conserved. By considering parental investments in fertility and education, the model demonstrates that as life expectancy rises, the optimal investment strategy shifts from Phase \one, where total population size is maximized and \(\lambda e_0\) is conserved, to Phase \two, where total productivity is maximized and $\lambda \exp(e_0 / 18)$ is conserved. Furthermore, the model suggests that lowering educational costs could help sustain fertility rates while preserving human capital development.

\section*{Results}
\subsection*{Two universal pathways in demographic transition}
\begin{figure*}[tb]
 \centering
  \includegraphics[width=\linewidth]{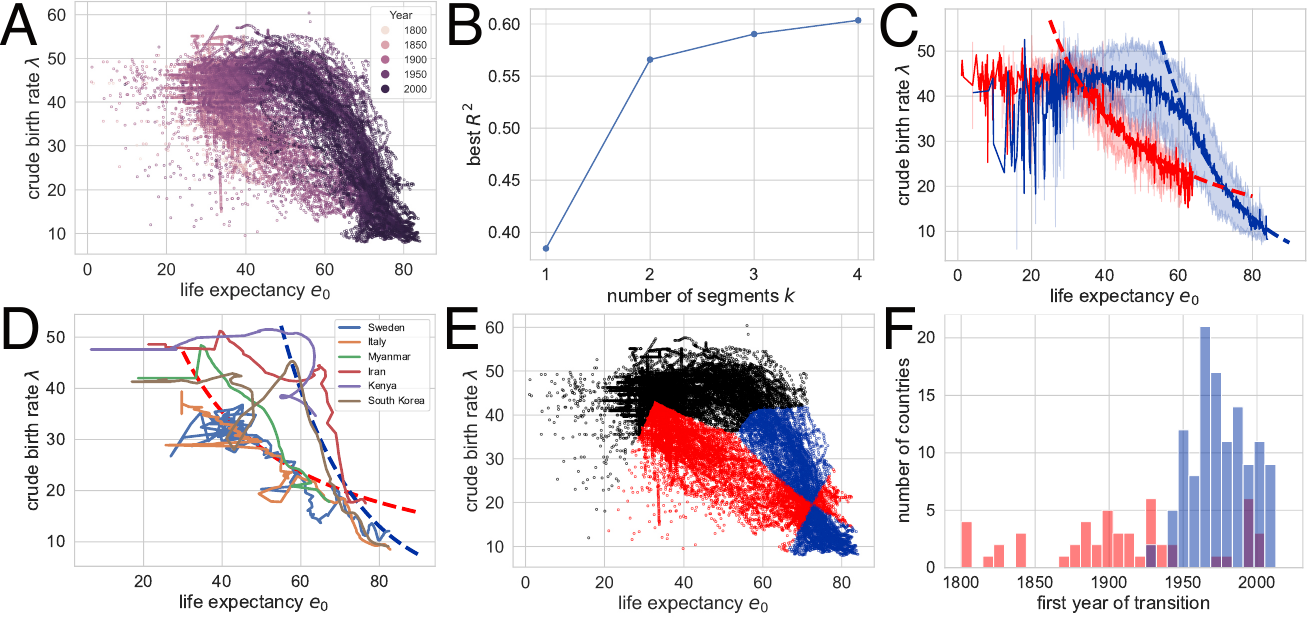}
\caption{
Relationship between the crude birth rate $\lambda$ and life expectancy at birth $e_0$. 
(A) Scatterplot of data from 195 countries (1800–2015), with colors indicating years.
(B) $R^2$ score as a function of the number of segments, where data points are partitioned, fitted independently, and the best score for each segmentation is shown.
(C) Two observed trends: pre-1930 (red) and post-1930 (blue). Solid lines represent average $\lambda$ for each $e_0$, with shaded areas indicating standard deviations. Dashed lines show the isoclines of $\lambda e_0$ (red) and $\lambda \exp(e_0 / 18)$ (blue).
(D) Sample pathways of countries, with dashed lines as isoclines from (C).
(E) Data classification into three categories: Phase \one\ (red), closer to the red dashed line; Phase \twoo\ (blue), closer to the blue dashed line; and pre-transition (black), closer to $\lambda = 43$.
(F) Onset years of each phase. Red and blue histograms indicate the number of countries entering Phase \one\ and Phase \twoo, respectively.
}
\label{fig:demographic_transition_path}
\end{figure*}
Figure \ref{fig:demographic_transition_path}(A) shows the global demographic trend where crude birth rate $\lambda$ decreases as life expectancy at birth $e_0$ increases, with most countries transitioning from the upper left (high $\lambda$, low $e_0$) to the lower right (low $\lambda$, high $e_0$).
A closer examination reveals that the data points are primarily concentrated along the edges of a triangle, with vertices approximately at $(\lambda, e_0) = (30, 50)$, $(60, 50)$, and $(80, 10)$. Older data (lighter colors) tend to cluster along the edge connecting $(30, 50)$ and $(80, 10)$, whereas more recent data (darker colors) are distributed along the edge between $(60, 50)$ and $(80, 10)$.

These findings suggest that the relationship between $\lambda$ and $e_0$ varies across distinct timespans. To identify the optimal number of timespans that best capture major trends, I varied the number of segments $k$ from 1 to 4 and systematically searched for the optimal partition. Each segment was independently fitted using either a power-law or exponential model, and the best $R^2$ values for each $k$ are shown in Fig. \ref{fig:demographic_transition_path}(B). While dividing the dataset into two segments significantly improves the fit compared to a single-segment model, further segmentation ($k=3$ or $k=4$) provides only marginal improvements. This suggests the existence of two primary phases in demographic transition. See the Materials and Methods section for details of the procedure.

By segregating the data into pre- and post-1930 segments, two distinct trends emerge. Fig. \ref{fig:demographic_transition_path}(C) shows that pre-1930 data cluster around the master curve $\lambda e_0 = 1416$ (red dashed line), while post-1930 data align with $\lambda \exp(e_0 / 18) = 1110$ (blue dashed line)\footnote{A similar trend is observed when setting the threshold year to either 1920 or 1940. Additionally, the blue dashed line can also be fitted with the curve $\lambda e_0^4 = 5.3 \times 10^8$.}.
Notably, the term $\lambda e_0 / 1000$ represents the population growth per generation, implying that the $\lambda e_0$ isocline indicates a state of steady population growth. Assuming a generational span of 25 years, $\log(1.416) / 25 \simeq 0.014$ suggests an annual population growth rate of approximately $1.4\%$. 
Conversely, the curve $\lambda \exp(e_0 / 18) = 1110$ reveals a sharper decline in $\lambda$ per unit increase in $e_0$. The divergence in conserved quantities highlights distinct mechanisms driving these transitions. 
Based on these findings, two phases of demographic transition are identified: Phase \one, which conserves $\lambda e_0$, and Phase \two, which conserves $\lambda \exp(e_0 / 18)$.

These differing trends cannot be attributed to the presence of an older population no longer contributing to fertility. Fig. S3 presents the relationship between $e_0$ and the rescaled birth rate, representing the number of births per working-age population. A similar pattern is observed in Fig. \ref{fig:demographic_transition_path}(C) and Fig. S3(B), reinforcing the conclusion that fundamentally distinct mechanisms drive fertility decline in the two phases.

Trajectories of individual countries follow one of the two master curves as shown in Fig. \ref{fig:demographic_transition_path}(D). Sweden (blue) and Italy (orange) experienced Phase \one\ before transitioning to Phase \two. Both initially followed the red curve downward, and upon reaching the intersection with the blue curve (approximately at $\lambda = 20$ and $e_0 = 70$), they transitioned along the blue curve, exhibiting a steeper decline. 
Notably, Sweden has been recognized as a country with recent improvements in fertility rates \cite{zaidi2017second}. However, Fig. \ref{fig:demographic_transition_path}(D) suggests that such improvements occur only along the blue master curve. In contrast, Myanmar's trajectory (green) closely follows the red curve, mirroring the pathway of Western countries with a delay of approximately a century.
The trajectories of Iran (red) and Kenya (purple) closely align with the blue curve, indicating that these countries did not experience Phase \one. Interestingly, South Korea (brown) initially underwent Phase \one, followed by Phase \two\ after a sudden increase in both fertility and longevity.
These countries were not selectively chosen to fit these patterns. The trajectories of all countries, shown in Figs. S4–S9, confirm the robustness of these trends. 
Thus, these two master curves are referred to as ``universal pathways'' in demographic transition.

Data points are classified into three classes: Phase \one\ (red), Phase \two\ (blue), and pre-transition (black) in Fig. \ref{fig:demographic_transition_path}(E). Fig. \ref{fig:demographic_transition_path}(F) illustrates that Phase \one\ can occur at any time, indicating its universal nature, whereas Phase \two\ is specific to the modern era. Many countries transitioned from Phase \one\ to Phase \two\ around 1950. Tables \ref{tab:DT_phase_excerpt} and S1 present the years in which each country experienced each phase.

\begin{table}[tb]
    \centering
    \caption{The year in which each country experienced each phase of demographic transition (excerpt). Yearly data up to 2015 were used for the analysis. The complete list is available in Table S1.}
    \label{tab:DT_phase_excerpt}
    \begin{tabular}{rll} 
Phase I	&	Country	&	Phase I\hspace{-1.2pt}I	\\
\hline
1822-1949	&	Sweden	&	1950-2015	\\
1896-1969	&	Italy	&	1970-2015	\\
1981-2015	&	Myanmar	&	---	\\
---	&	Iran	&	1986-2008	\\
---	&	Kenya	&	2005-2015	\\
1949-1952	&	South Korea	&	1963-2015
    \end{tabular}
\end{table}

\begin{figure*}[tb]
 \centering
  \includegraphics[width=.9\linewidth]{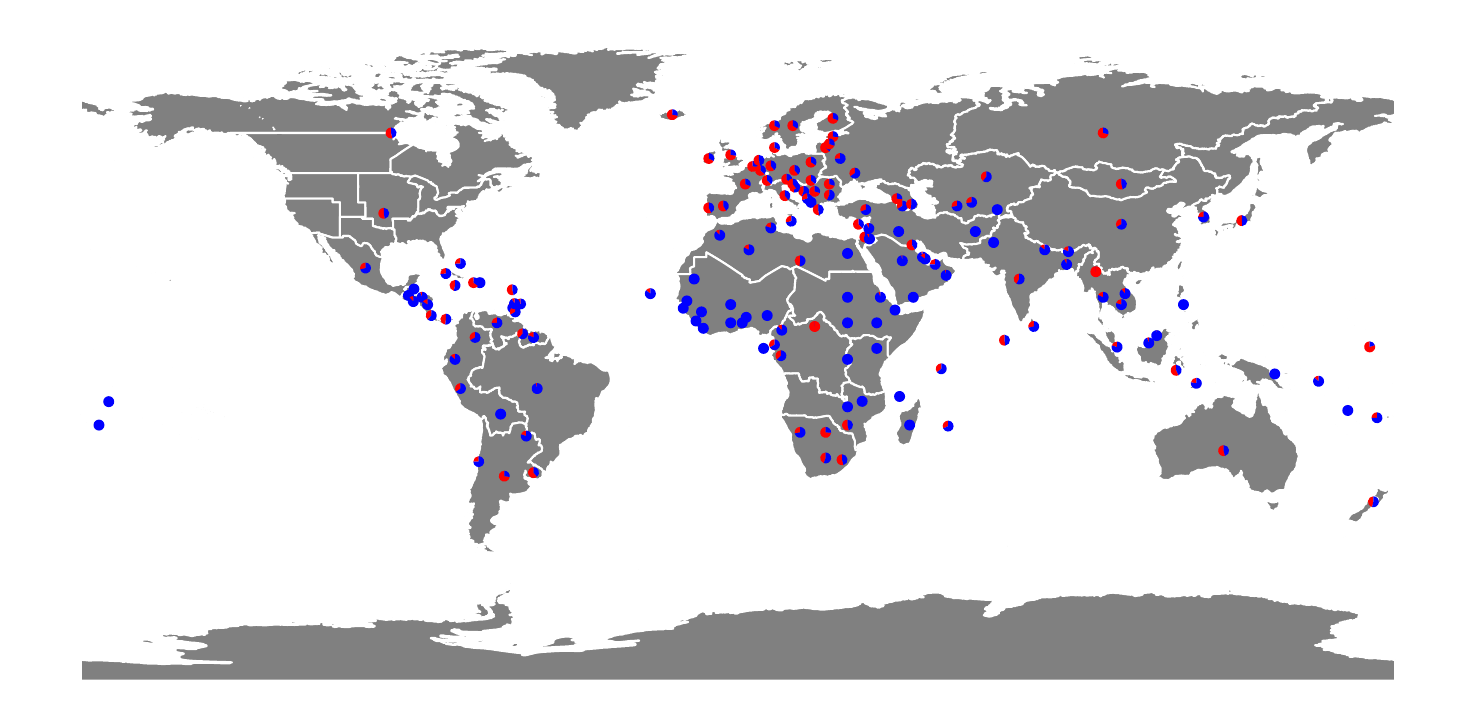}
\caption{
The distribution of countries experiencing Phases \one\ and \twoo. The pie charts illustrate the proportion of years spent in Phase \one\ (red) and Phase \twoo\ (blue).
}
\label{fig:demographic_transition_worldmap}
\end{figure*}

\begin{figure*}[tb]
 \centering
  \includegraphics[width=\linewidth]{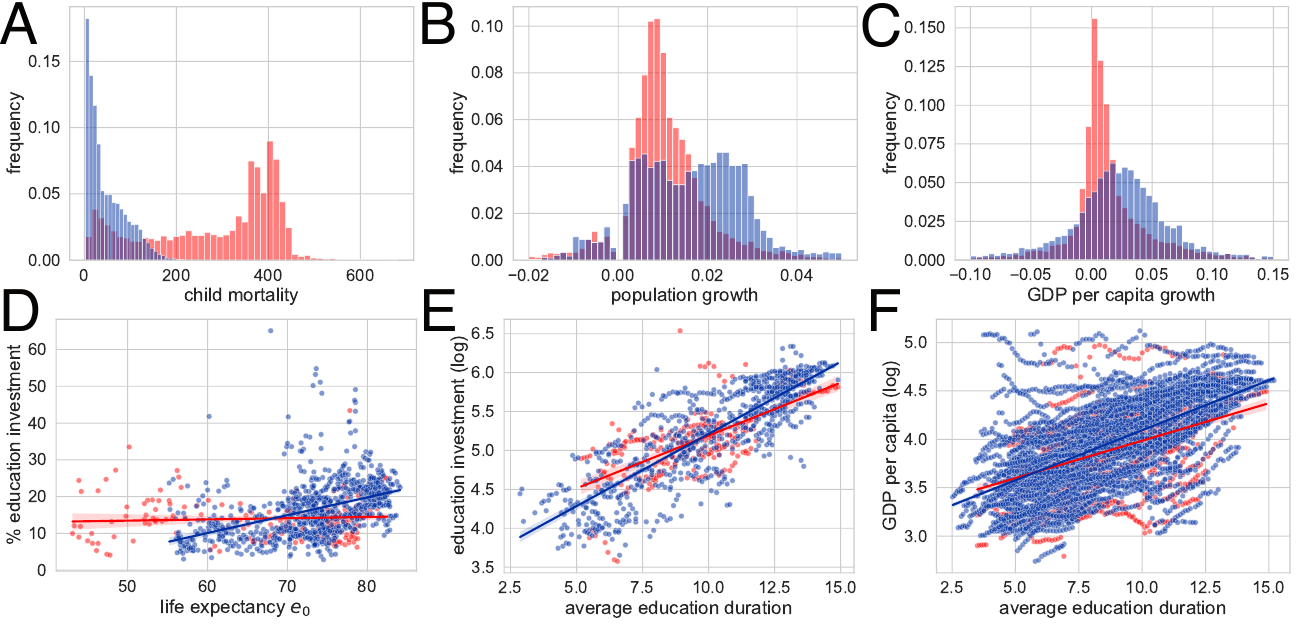}
  \caption{
Indices characterizing the two phases of demographic transition. The histograms present (A) child mortality rate (average number of deaths of children under five per 1,000 births), (B) population growth rate, and (C) GDP per capita growth rate. 
Panels (D)–(F) illustrate relationships between demographic and educational indicators:  
(D) Life expectancy at birth ($e_0$) versus education investment per student relative to GDP per capita.  
(E) Average educational duration versus education investment per student (log scale, USD).  
(F) Average educational duration versus GDP per capita, representing the return on educational investment.  
Red and blue indicate data corresponding to Phases \one\ and \twoo, respectively. The red and blue lines in (D)–(F) represent linear regression results.
}
\label{fig:demographic_transition_indices}
\end{figure*}

Figure \ref{fig:demographic_transition_worldmap} illustrates the geographical distribution of phases. Phase \one\ is prevalent—though not exclusive to—Western Europe, North America, and East Asia, with most transitioning to Phase \two\ after 1950. In contrast, the majority of countries in Africa, South America, and South Asia have experienced only Phase \two. 
Phase \one\ is typically observed in countries where fertility declined until the mid-20th century. However, some countries, such as Myanmar (Fig. \ref{fig:demographic_transition_path}(D)), experienced it in the latter half of the 20th century.

\subsection*{Two phases of demographic transition}
Figure \ref{fig:demographic_transition_indices} shows the distinct features of the two phases. Phase \one\ is characterized by high child mortality and a steady but low rate of population growth, as shown in Fig. \ref{fig:demographic_transition_indices}(A, B). In contrast, Fig. \ref{fig:demographic_transition_indices}(A, C) shows that Phase \two\ is marked by lower child mortality and steady growth in GDP per capita rather than in population size. 
Moreover, Fig. S10 shows that although $e_0$ increases at a similar pace in both phases, the decrement of $\lambda$ is larger in Phase \two.

Educational investment, particularly when measured as a percentage of GDP per capita, remains relatively low and depends minimally on $e_0$ in Phase \one, while increases exponentially with $e_0$ in Phase \two, as shown in Fig. \ref{fig:demographic_transition_indices}(D) and Fig. S11. 
As many countries shift from Phase \one\ to Phase \two\ around $e_0 = 70$ (i.e. the intersection of two curves), a sharp increase in educational investment follows this shift. 
The correlation between $e_0$ and education expenditure as a percentage of GDP per capita is $0.08$ for Phase \one\ and $0.43$ for Phase \two. Similarly, the correlation between $e_0$ and the logarithm of education investment per student is $0.56$ for Phase \one\ and $0.86$ for Phase \two\ (Fig. S11). 
Additionally, Fig. \ref{fig:demographic_transition_indices} (E, F) indicates that, in both phases, educational investment and GDP per capita (i.e., its return) increase exponentially with education duration.

These trends align with the broader societal shift toward ``higher-order needs,'' such as self-realization, expressive work, and educational values, which characterize the second demographic transition \cite{maslow1954motivation, lesthaeghe2014second, inglehart2018culture}. They also reflect the ``quantity-quality trade-off'' in economics and human behavioral ecology, wherein fertility is balanced against investment in human capital through education \cite{hanushek1992trade, fernihough2017human, kaplan1996theory, colleran2015fertility}.

Therefore, Phase \one, the movement along the red curve in Fig. \ref{fig:demographic_transition_path}(C), corresponds to the first demographic transition, which is typically characterized by a decline in fertility to the replacement level (i.e., two children per woman).
As life expectancy increases, a greater proportion of women live beyond age 50, allowing the total fertility rate (TFR)—which measures the hypothetical number of children a woman would have if she survived to age 50—to align more closely with the actual average number of children per woman. 
Since Phase \one\ is characterized by slow but steady population growth with a constant $\lambda e_0$, TFR gradually declines to $2$.

Still, the observed trends in Phase \two\ highlight the limitations of categorizing demographic transitions solely based on TFR. In Fig. \ref{fig:demographic_transition_path}(C), the segment of the blue curve of Phase \two\ below the red curve of Phase \one\ signifies a fertility decline below the replacement level, defining the second demographic transition. However, states in Phase \two\ that lie above the red curve, often observed in developing countries, present a puzzle. 
Traditional classifications attribute these states to the first demographic transition, as TFR exceeds 2. Yet, the present findings suggest that these states align more closely with the second demographic transition in Western countries, as they follow the same universal pathway.

These two phases are also consistent with the growth phases proposed in the unified growth theory \cite{galor2011unified}. Phase \one\ corresponds to the Malthusian growth phase, characterized by steady population growth, while Phase \two\ corresponds to the modern growth phase, which is marked by sustained growth in GDP per capita and technological levels.

\subsection*{Origin of universal pathways}
Now, let us investigate the origin of the two universal pathways. As described in the Materials and Methods section, $\lambda e_0$ remains conserved under conditions of steady population growth \cite{preston2000demography}. 
Thus, Phase \one\ follows the Malthusian growth model, where fertility decline is driven by population pressure. This scenario can be interpreted as the optimization of the total quantity of people. In contrast, when the focus shifts to optimizing the total quality of the population, a different conservation law will emerge, whereby living twice as long has distinct effects compared to having twice the population.

I introduce a simple model of parental investment in children to elucidate the transition between Phases \one\ and \two. In this model, life expectancy, $e_0$, is assumed to be given, as mortality decline is often regarded as an exogenous factor driving fertility decline \cite{dyson2010population, canning2011causes}. It is assumed that education imposes costs on parental resources, reducing fertility, as well as on children's time, limiting their participation in productive activities \cite{hanushek1992trade, hedges2018trade}.
Then, the number of children, $\lambda$, and the fraction of lifetime allocated to education, $p$, are optimized. Children devote a duration of $pe_0$ to education, while the remaining $(1-p)e_0$ is allocated to productive activities. A unit cost is incurred annually for a child’s survival. Consequently, if $\lambda$ children live for $e_0$ years, the living cost is $\lambda e_0$. 

Additionally, parents may invest in their children's education to improve their productivity. It is assumed that both the cost and return of education increase exponentially with the duration of education, following empirical observations in Fig. \ref{fig:demographic_transition_indices}(E, F). Such a nonlinear increase is reasonable, considering that higher education often requires more specialized personnel and materials. 

The educational cost is modeled as $c\exp(\beta pe_0) - c$, where $\beta$ determines the degree of nonlinearity, and $c$ represents the unit educational cost. Similarly, production efficiency increases by $\alpha \exp(\beta pe_0) - \alpha$, where $\alpha$ is the unit increment in efficiency. The terms $c$ and $\alpha$ are subtracted to ensure that both cost and efficiency increment are zero when $p = 0$.

The number of children, $\lambda$, and the fraction of education, $p$, are determined by solving the following optimization problem. The total cost, defined as the sum of living and education costs, is subject to the constraint:
\begin{equation}
    \lambda(e_0 + c \exp(\beta p e_0) - c) \leq 1.
\end{equation}
Under this constraint, $\lambda$ and $p$ are optimized to maximize the total productivity of children:
\begin{equation}
    \max_{\lambda,\ p} \lambda (1 - p)e_0(1 + \alpha \exp(\beta p e_0) - \alpha),
\end{equation}
where $\lambda$ children allocate $(1 - p)e_0$ years to productive activities, with a productivity of $1 + \alpha \exp(\beta p e_0) - \alpha$. While the values of $\alpha$, $\beta$, and $c$ may differ across policies, institutions, and industrial structures, I first analyze the dependence of $\lambda$ and $p$ on $e_0$ by holding these parameters constant.

\begin{figure}[tb]
 \centering
  \includegraphics[width=\linewidth]{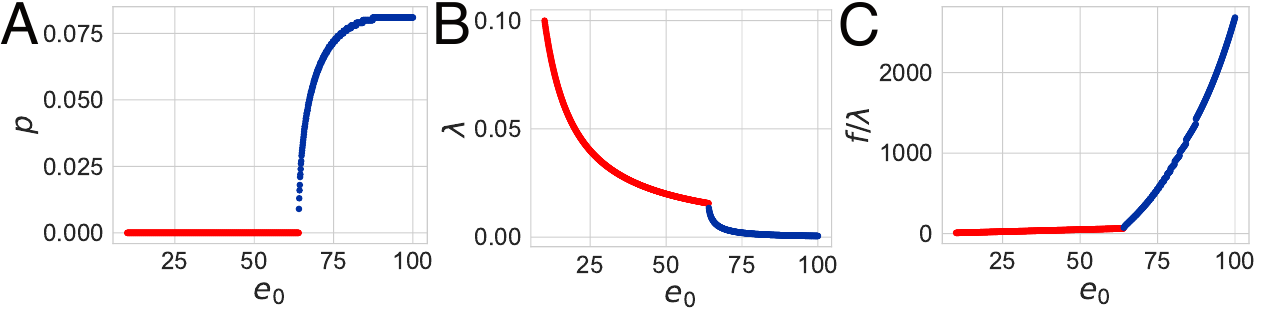}
  \caption{
Results of numerical calculations. (A) The optimal fraction of educational duration, $p$, (B) the optimal fertility rate, $\lambda$, and (C) per capita productivity, $f / \lambda$, as functions of life expectancy, $e_0$. 
In this calculation, the parameters are set to $\alpha = 0.5$, $\beta = 0.5$, and $c = 30$. Points corresponding to Phase \one, where $p = 0$, are plotted in red, while those for Phase \twoo, where $p > 0$, are plotted in blue.
}
\label{fig:demographic_transition_numerical}
\end{figure}

It is important to note that the optimal $\lambda$ is proportional to the total cost, implying that the optimal $p$ remains independent of its value. Without loss of generality, the total cost can be normalized to 1, resulting in relatively small values of $\lambda$ in Fig. \ref{fig:demographic_transition_numerical} compared to the observed crude birth rate. The parameters used in the model are summarized in Table \ref{table:DT_params}. Further details of the model can be found in the Materials and Methods section.

\begin{table}[tb]
\caption{Model parameters. In this model, $e_0$, $\alpha$, $\beta$, and $c$ are predetermined, while $\lambda$ and $p$ are optimized.}
  \label{table:DT_params}
   \centering
    \begin{tabular}{ll} 
sign & explanation \\
\hline
$e_0$ & Life expectancy of children \\
$\alpha$ & Increment of productivity by education \\
$\beta$ & Nonlinearity in educational investment \\
$c$ & Increment of cost for education\\
$\lambda$ & The (relative) number of births \\
$p$ & The fraction of education duration 
    \end{tabular}
\end{table}

Numerical calculations in Fig. \ref{fig:demographic_transition_numerical} reveal that as life expectancy $e_0$ increases, educational investment for children becomes suddenly advantageous, delineating two distinct phases. Fig. \ref{fig:demographic_transition_numerical}(A) shows a sudden rise in the optimal $p$ at a specific $e_0$ threshold. For small $e_0$, the optimal $p$ equals zero, conserving $\lambda e_0$. This phase corresponds to Phase \one, where the strategy prioritizes maximizing population size without educational investment. As $e_0$ increases, longer lifespans enable extended productive work, making educational investment more feasible. Consequently, the optimal $p$ becomes positive and remains relatively stable, leading to a gradual increase in the optimal educational duration, $pe_0$. In the phase where $p > 0$, the quantity $\lambda (c\exp(\beta p^\ast e_0) + e_0 - c) \simeq c\lambda \exp(\beta p^\ast e_0)$ is conserved, characterizing Phase \two. The observed conservation of $\lambda \exp(e_0 / 18)$ implies that $\beta p^\ast \simeq 1 / 18$. This shift aligns with Fig. \ref{fig:demographic_transition_indices}(D), where educational investment transitions from being minimal and $e_0$-independent in Phase \one\ to increasing with $e_0$ in Phase \two.

The optimal fertility in Fig. \ref{fig:demographic_transition_numerical}(B) illustrates a seamless and spontaneous transition from Phase \one\ to Phase \two\ as a function of $e_0$, resembling the trajectories of many Western countries (e.g., Italy and Sweden in Fig. \ref{fig:demographic_transition_path}(D)). 
By contrast, extrapolating the right-hand branch where $p > 0$ reproduces the blue master curve in Fig. \ref{fig:demographic_transition_path}(C). 
The trajectories of many developing countries, which either did not experience Phase \one\ or exited it midway, follow this extrapolated blue curve, suggesting that their demographic dynamics were influenced by cultural transmission or Westernization \cite{lesthaeghe2010unfolding, amin2002spatial, colleran2016cultural}. 
In other words, their demographic transition may have resulted from the adoption of institutional parameters ($\alpha, \beta$, and $c$) and/or behavioral strategies prioritizing child education over fertility, imported from Western countries in Phase \two.

Finally, Fig. \ref{fig:demographic_transition_numerical}(C) illustrates per capita productivity, which serves as an indicator of GDP per capita. It remains nearly constant within Phase \one\ and increases steadily with $e_0$ during Phase \two, consistent with the trends observed in Fig. \ref{fig:demographic_transition_indices}(C).

\begin{figure}[tb]
 \centering
  \includegraphics[width=\linewidth]{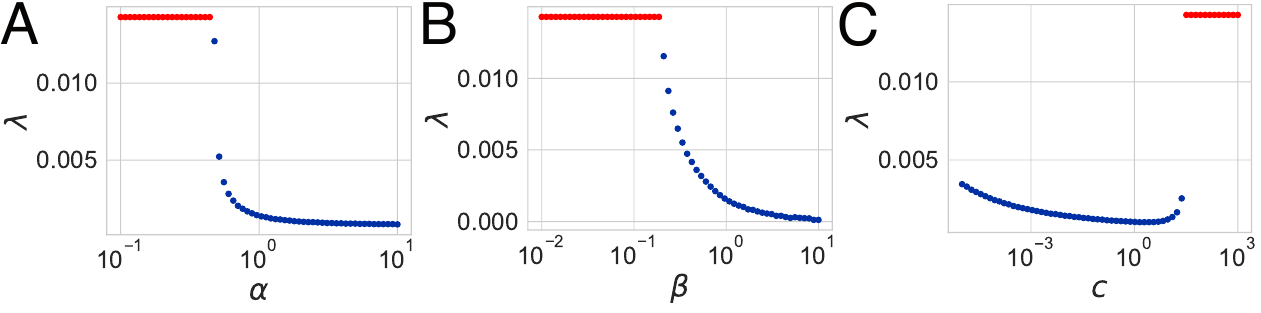}
  \caption{
Dependence of the optimal fertility, $\lambda$, on (A) the educational efficiency, $\alpha$, (B) the nonlinearity in educational investment, $\beta$, and (C) the educational cost, $c$. 
Unless otherwise specified, the parameters are set to $\alpha = 0.5$, $\beta = 0.5$, $c = 30$, and $e_0 = 70$. Points corresponding to Phase \one, where $p = 0$, are shown in red, while those for Phase \twoo, where $p > 0$, are shown in blue. The downward arrow in (C) represents the proposed policy intervention to enhance fertility.
}
\label{fig:demographic_transition_policy}
\end{figure}

Next, the impact of various policies on optimal fertility is examined by analyzing its dependence on parameters. By fixing $e_0$ at 70 and varying the values of $\alpha$, $\beta$, and $c$, the optimal value of $\lambda$ is determined. 
Fig. \ref{fig:demographic_transition_policy} shows that fertility is highest when the increment in production efficiency, $\alpha$, and the nonlinearity in educational investment, $\beta$, are small, while the educational cost, $c$, is high. However, this region corresponds to Phase \one, where $p = 0$. 
This scenario, characterized by ineffective educational investment, represents an undesirable outcome from a policy perspective.

Conversely, Fig. \ref{fig:demographic_transition_policy}(C) indicates that a significant reduction in the educational cost, $c$, leads to an increase in fertility (as indicated by the arrow), while remaining in Phase \two. 
This suggests that making education more affordable allows resources to be reallocated toward reproduction without compromising educational quality. Such a scenario is favorable, as it enables both the maintenance of a healthy population growth rate and sustained investment in education. 
Moreover, global data support this observation, as shown in Fig. S12, where lower values of $c$ are associated with higher fertility rates.

\section*{Discussion}
\subsection*{Universal mechanisms of demographic transition}
By examining the relationship between the crude birth rate, $\lambda$, and life expectancy at birth, $e_0$, two universal pathways of demographic transition have been identified, characterized by the conservation of $\lambda e_0$ and $\lambda \exp(e_0 / 18)$, respectively. 
These pathways define two distinct phases: Phase \one, which broadly corresponds to the first demographic transition observed in Western countries and the Malthusian growth phase, and Phase \two, which aligns with the second demographic transition and modern growth phase\footnote{Although some studies associate the modern growth phase with the first demographic transition \cite{zaidi2017second}, the present findings suggest that it more accurately coincides with the second.} \cite{lesthaeghe2014second, zaidi2017second, galor2011unified}.

Phase \one\ has occurred at any time over the past 200 years, whereas Phase \two\ is unique to the modern era, emerging predominantly after 1950. 
Phase \one\ was prevalent in Western Europe, North America, and East Asia, characterized by high child mortality and steady population growth. Most of these regions transitioned to Phase \two\ around 1950. 
In contrast, most countries in Africa, South America, and South Asia experienced only Phase \two, marked by low child mortality and steady growth in GDP per capita. This raises questions about the applicability of the first and second demographic transition frameworks outside of Western contexts.

This study highlights the universality of these phases by utilizing comprehensive global data. Interestingly, vertebrates also exhibit a linear relationship between fertility and mortality, conserving $\lambda e_0$, as in Phase \one\ \cite{ricklefs2010life}. This finding suggests that Phase \one\ represents a universal pattern in nature, whereas Phase \two\ appears to be a uniquely human phenomenon.

A simple model was then proposed by considering the trade-off between reproduction and education. Numerical analysis demonstrated the transition between Phase \one, where reproduction is prioritized with minimal investment in education, and Phase \two, where education is emphasized, even at the cost of reduced fertility.

Phase \one\ emerges when life expectancy is high and affordable education significantly enhances productivity. In this phase, the total population size is maximized, and $\lambda e_0$ is conserved. 
Phase \two\ occurs under opposite conditions, where total productivity is maximized, and $\lambda \exp(\beta p^\ast e_0)$ is conserved. Here, $\beta$ represents the nonlinearity in educational investment, and $p^\ast$ denotes the optimal fraction of educational duration per lifetime. This phase transition corresponds to the shift from the first to the second demographic transition and from the Malthusian to the modern growth phase \cite{lesthaeghe2014second, galor2011unified}. 

The parameter dependence of optimal fertility has significant policy implications. It is proposed that reducing educational costs ($c$) facilitates the reallocation of resources toward reproduction while maintaining adequate educational standards. By alleviating the financial burden of education, policymakers can foster an environment in which fertility recovers without compromising human capital development.

The theoretical results align with sociological observations, highlighting the critical role of education in demographic transition. Modernization has facilitated greater social mobility through mass education at lower costs, supported by a meritocratic ideology \cite{breen2007explaining, squicciarini2016knowledge, gil2017intergenerational}. This institutional transformation incentivizes educational investment, effectively increasing educational efficiency ($\alpha$) and reducing educational cost ($c$), which are required for the emergence of Phase \two.

\subsection*{Connection to demographic theories}
This study establishes macroscopic laws by identifying conserved quantities, a method commonly employed in physics. It provides the necessary conditions that any explanatory model of demographic transition must satisfy. Although explanations have been proposed from various disciplines \cite{galor2011unified, mace2000evolutionary, ihara2004cultural, cervellati2005human}, models that fail to yield the conservation of $\lambda e_0$ or $\lambda \exp(e_0 / 18)$ can be considered insufficient. Therefore, the present findings are essential for advancing the theoretical understanding of demographic transition.


Previous explanations of fertility decline can be broadly classified into three categories based on their primary focus: (i) population pressure \cite{malthus1798essay}, (ii) the trade-off between education and fertility \cite{hanushek1992trade, bleakley2009chronic, galor2011unified, canning2011causes, becker2010trade, werding2014children, fernihough2017human, kaplan1996theory, mace2000evolutionary, lawson2011parental, shenk2009testing}, and (iii) the cultural transmission of behavioral strategies \cite{amin2002spatial, murphy2017economization, mulder1998demographic, ihara2004cultural, richerson2008not, newson2007influences}.
The present framework integrates these perspectives by identifying (i) as the mechanism driving Phase \one, (ii) as the driver of Phase \two, and (iii) as the key factor facilitating the transition from Phase \one\ to Phase \two.

The recent rapid fertility decline—particularly in regions that have not experienced Phase \one—has been linked to Westernization \cite{kirk1996demographic, murphy2017economization}. 
The present findings suggest that Westernization may have facilitated the cultural transmission of parental investment strategies prioritizing education, or the adoption of institutional parameters such as educational efficiency ($\alpha$), the nonlinearity of educational investment ($\beta$), and educational cost ($c$), thereby accelerating the transition to Phase \two. The transmission of reproductive behavior has also been highlighted in previous studies \cite{amin2002spatial, Borenstein2006cultural}. 
Such transmission offers a plausible explanation for the demographic trajectories of many developing countries, where Phase \one\ was either not observed or was exited midway.

Educational investment as a driver of fertility decline in Phase \two\ offers additional insights. It is indicated that uncertainty in life expectancy decreases as child mortality declines, a trend observed for Phase \two\ in Fig. \ref{fig:demographic_transition_indices}(A) \cite{aburto2020dynamics}. This reduction in uncertainty has been proposed to drive fertility decline by lowering the risk of lineage extinction \cite{leslie2002demographic}. 
The present theory further suggests that in environments characterized by lower child mortality and more predictable life expectancy, the incentive to invest in education intensifies, accelerating the transition to Phase \two. Moreover, extended educational attainment and subsequent labor force participation, particularly among women, have been associated with delayed first births, contributing to fertility decline \cite{snopkowski2016pathways}.


The present analysis elucidates the significance of the conserved quantities \(\lambda e_0\) and \(\lambda \exp(\beta p^\ast e_0)\) and explains their underlying mechanisms, although their specific values remain unclear. As shown in Fig.~\ref{fig:demographic_transition_path}(C), \(\lambda e_0 \approx 1400\), indicating an annual population growth rate of about 1.4\%. In the model, this would require the cost constraint to rise at the same rate in most countries, yet this rate cannot be derived theoretically. Likewise, the conservation of \(\lambda \exp(e_0 / 18) \approx 1100\), including why \(\beta p^\ast = 1/18\) and how $1100$ arises, remains insufficiently explained.
The clustering of demographic trajectories around these universal pathways (Fig.~\ref{fig:demographic_transition_path}(C)) may reflect a common constraint on parameter values, potentially rooted in biological or sociocultural factors. Nevertheless, determining the precise numerical values of these conserved quantities remains a key limitation of this study.

Additionally, several limitations should be acknowledged. First, the current model focuses only on the trade-off between reproduction and education, neglecting other trade-offs, such as those involving parents' career advancement and leisure \cite{hakim2003new, vitali2009preference, galindev2011leisure, kato2021low, debnath2022inter}. Further empirical and theoretical studies are needed to assess whether these trade-offs contribute to a third phase of demographic transition or are embedded within Phase \two.
Second, it remains unclear why some countries, such as South Korea (Fig. \ref{fig:demographic_transition_path}(D)), transitioned from Phase \one\ to Phase \two\ midway, while others, like Italy and Sweden, stayed in Phase \one\ until reaching the intersection of the two pathways. Countries transitioning midway exhibited increases in both fertility and longevity. Identifying the factors behind these differing trajectories is a critical area for future research.

In conclusion, this study identifies two universal pathways in demographic transition by analyzing the relationship between the crude birth rate, $\lambda$, and life expectancy at birth, $e_0$, across countries worldwide. These pathways, characterized by the conserved quantities $\lambda e_0$ and $\lambda \exp(e_0 / 18)$, correspond to phases where reproduction or education is prioritized. 
This work provides a novel perspective on demography by uncovering universal patterns from global data and integrating them with a robust theoretical framework.

\section*{Materials and Methods}
\subsection*{Data}
The dataset analyzed in this study was obtained from Gapminder (\url{https://www.gapminder.org/data/}). All available data from the database were included in the analysis, covering 195 countries from 1800 to 2015, although some data points are missing. 
While national boundaries may have changed over time, the dataset provides data for regions corresponding to present-day country boundaries.
The source codes of this study are available at: \url{https://github.com/KenjiItao/DemographicTransition.git}.

\subsection*{Optimization of the number of segments}
The relationship between the crude birth rate, $\lambda$, and life expectancy at birth, $e_0$, was analyzed using a piecewise regression approach. The dataset was divided into temporal segments $(1800-t_1, t_1-t_2, \cdots, t_{k-1}-2015)$ defined by threshold years $(t_1, t_2, \cdots, t_{k-1})$, with each segment independently fitted using either a power-law or exponential model. 

The power-law model is expressed as:
$\lambda = C/e_0^\alpha,$
where $C$ and $\alpha$ are parameters, while the exponential model is given by:
$\lambda = C / \exp(e_0 / \beta),$
where $C$ and $\beta$ are parameters. 

For each segment, the model with the lowest sum of squared residuals (SSE) was selected. The threshold years $t_1, t_2, \ldots, t_{k-1}$ were selected to maximize the overall coefficient of determination (\(R^2\)), calculated as:
$R^2 = 1 - \text{SSE}_{\text{total}}/\text{TSS}_{\text{total}},$
where \(\text{TSS}_{\text{total}}\) represents the total sum of squared deviations of \(\lambda\) from its mean.

\subsection*{Derivation of the Conservation of $\lambda e_0$}
The increase in population per 1,000 individuals is represented by the crude birth rate, $\lambda$, while the decrease is given by the crude death rate, which, under steady-state conditions, equals $1000 / e_0$ \cite{preston2000demography}. 
When the total population size remains constant, the birth and death rates are balanced, leading to the equation: $\lambda = 1000 / e_0$
which results in the conservation law: $\lambda e_0 = 1000.$
Similarly, the term $\lambda e_0 / 1000$ serves as a measure of population growth.

\subsection*{Model}
The model can be expressed as the following optimization problem. Subject to the cost constraint:
\begin{equation}
    \lambda(e_0 + c \exp(\beta p e_0) - c) \leq 1,
\end{equation}
$\lambda$ and $p$ are determined to maximize:
\begin{equation}
    \max_{\lambda,\ p} \lambda (1 - p)e_0(1 + \alpha \exp(\beta p e_0) - \alpha).
\end{equation}
As total productivity is proportional to $\lambda$, $\lambda(e_0 + c \exp(\beta p e_0) - c) = 1$, when it is maximized. Then, the optimal $\lambda$ satisfies:
\begin{equation}
    \lambda = \frac{1}{e_0 + c\exp(\beta pe_0) - c}.
\end{equation}
Thus, the optimization problem is reformulated as:
\begin{equation}
    \max_p f(p) = \frac{(1 - p)e_0(1 + \alpha \exp(\beta pe_0) - \alpha)}{e_0 + c\exp(\beta pe_0) - c}.
\end{equation}
Although the derivative $f'(p) = 0$ cannot be solved analytically, numerical calculations in Fig. \ref{fig:demographic_transition_numerical}(A), indicate that the optimal $p$ exhibits a sudden increase at a certain threshold as $e_0$ increases. 

\subsection*{Acknowledgement}
The author thanks Kunihiko Kaneko, Koji Hukushima, Bret Beheim, Heidi Colleran, Yamato Arai, and Ryohei Mogi for stimulating discussions.
This research was supported by JSPS KAKENHI Grant number JP21J21565 and Special Postdoctoral Researcher Program in RIKEN Project code 202401061006.

\bibliographystyle{unsrt}

\supplementaryfigures
\supplementarytables
\section*{Supplementary Figure}
\begin{figure}[H]
 \centering
  \includegraphics[width=.5\linewidth]{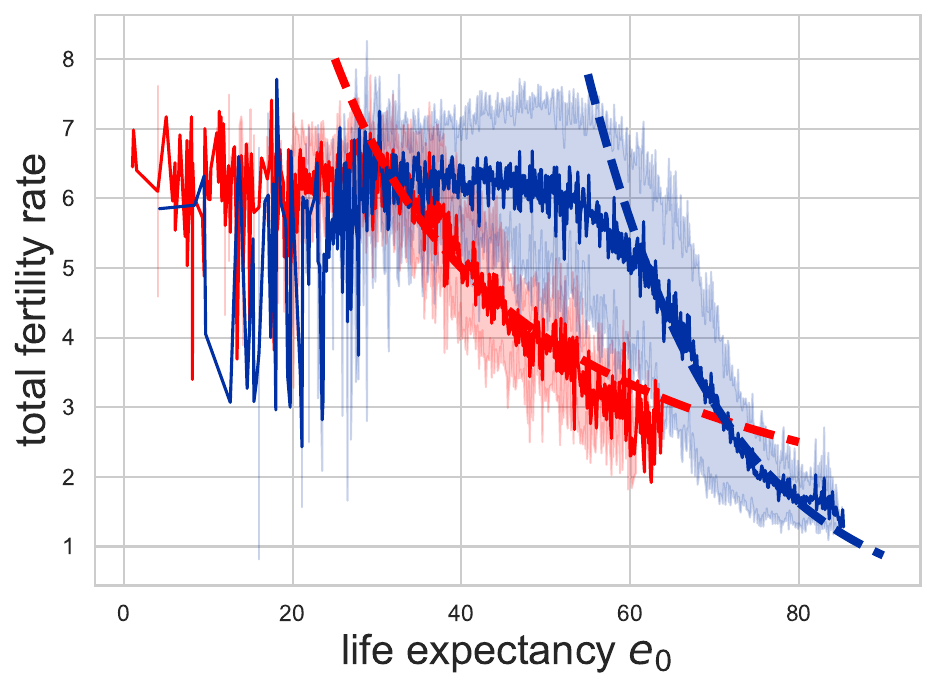}
\caption{
The relationship between the total fertility rate (TFR) and life expectancy, $e_0$. The analysis reveals that the observed trends remain consistent even when TFR is used in place of the crude birth rate, $\lambda$.
}
\label{fig:TFR_e0}
\end{figure}

\begin{figure}[H]
 \centering
  \includegraphics[width=\linewidth]{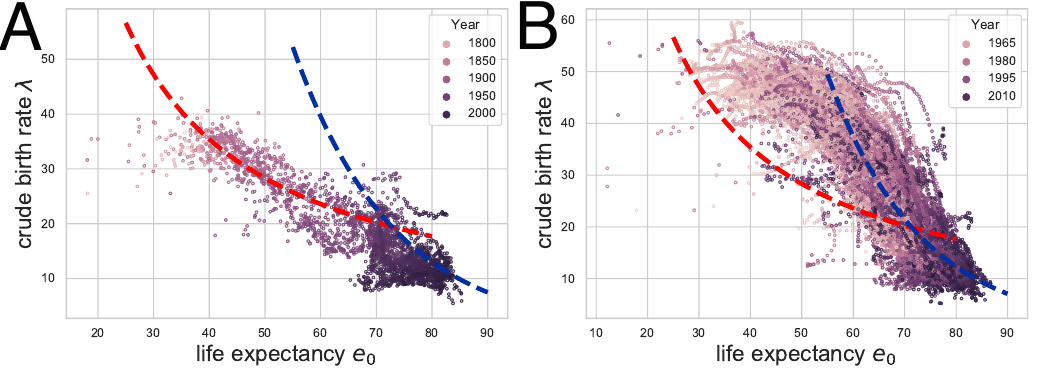}
  \caption{
The relationship between the crude birth rate, $\lambda$, and life expectancy, $e_0$, based on data from the Human Mortality Database (HMD) (A) and the United Nations Statistics Division (UN) (B). 
HMD provides extensive data primarily for Western countries, which are predominantly in Phase \one. In contrast, the UN dataset includes more recent data, capturing countries primarily in Phase \two.
}
\label{fig:UN_HMD}
\end{figure}

\begin{figure}[H]
 \centering
  \includegraphics[width=\linewidth]{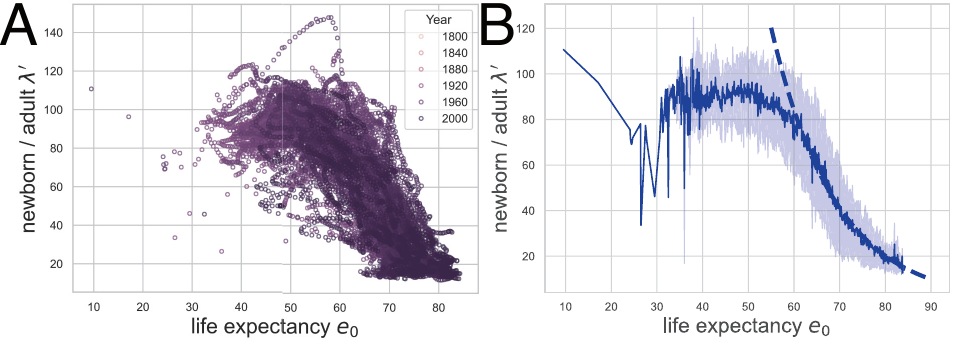}
\caption{
The relationship between the rescaled birth rate $\lambda'$ and life expectancy $e_0$. $\lambda'$ is defined as the ratio of the crude birth rate $\lambda$ to the working-age (15-60) population ratio, to exclude the decline in fertility due to life after menopause. Only post-1950 data are available, so those for phase \two\ are shown. (A) Scatterplot of data for 195 countries from 1950 to 2020. Colors represent the year. (B) The trend is similar to the blue curve in Fig. 1(B). The curve shows the isocline of $\lambda' \exp(e_0 / 14) = 4453$.
}
\label{fig:demographic_transition_appendix}
\end{figure}

\begin{figure}[H]
\centering
\includegraphics[width=.95\textwidth]{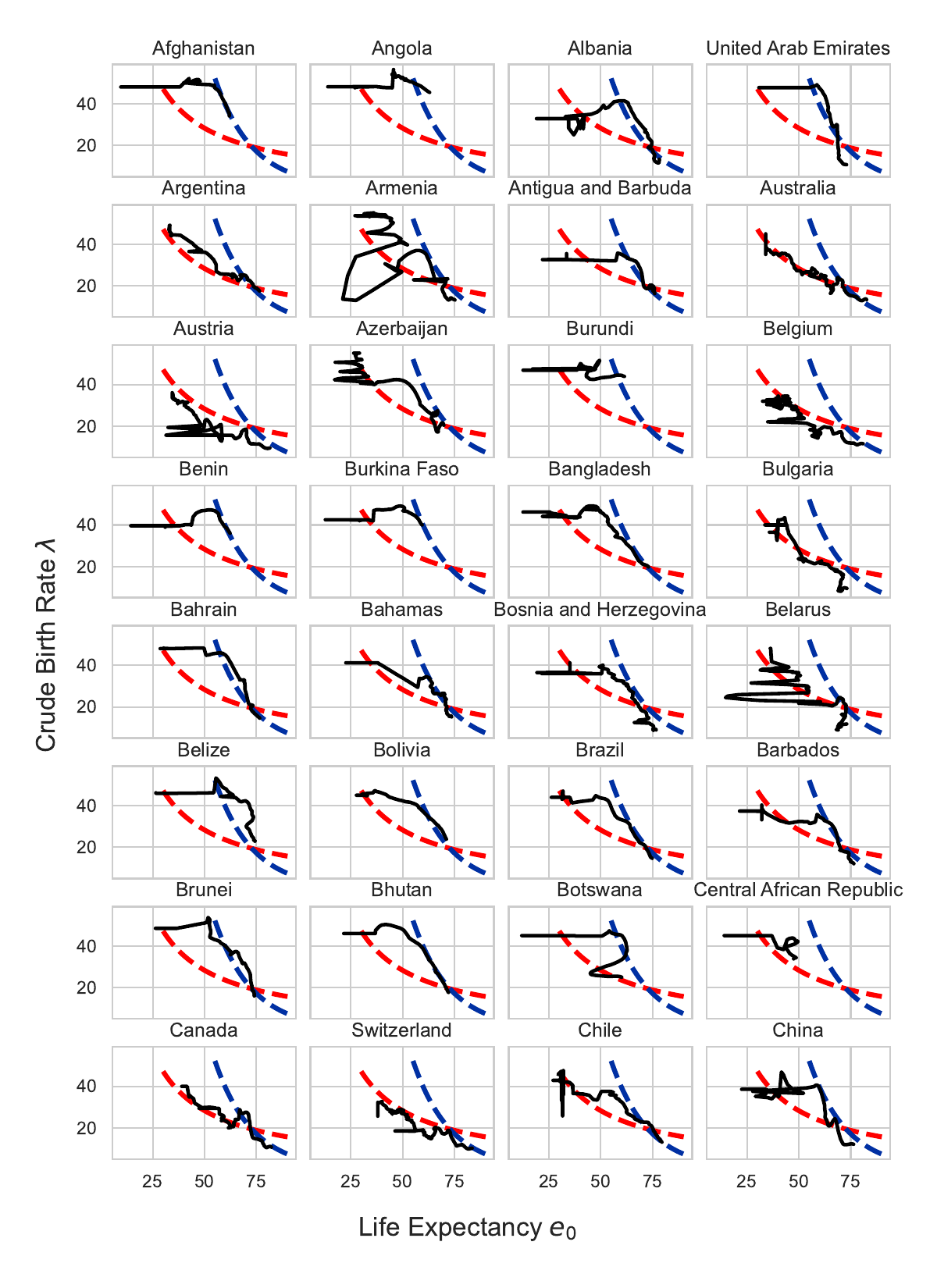}
\caption{Pathways of countries, with two universal pathways as dashed lines.
}
\label{fig:pathway_1}
\end{figure}

\begin{figure}[H]
\centering
\includegraphics[width=.95\textwidth]{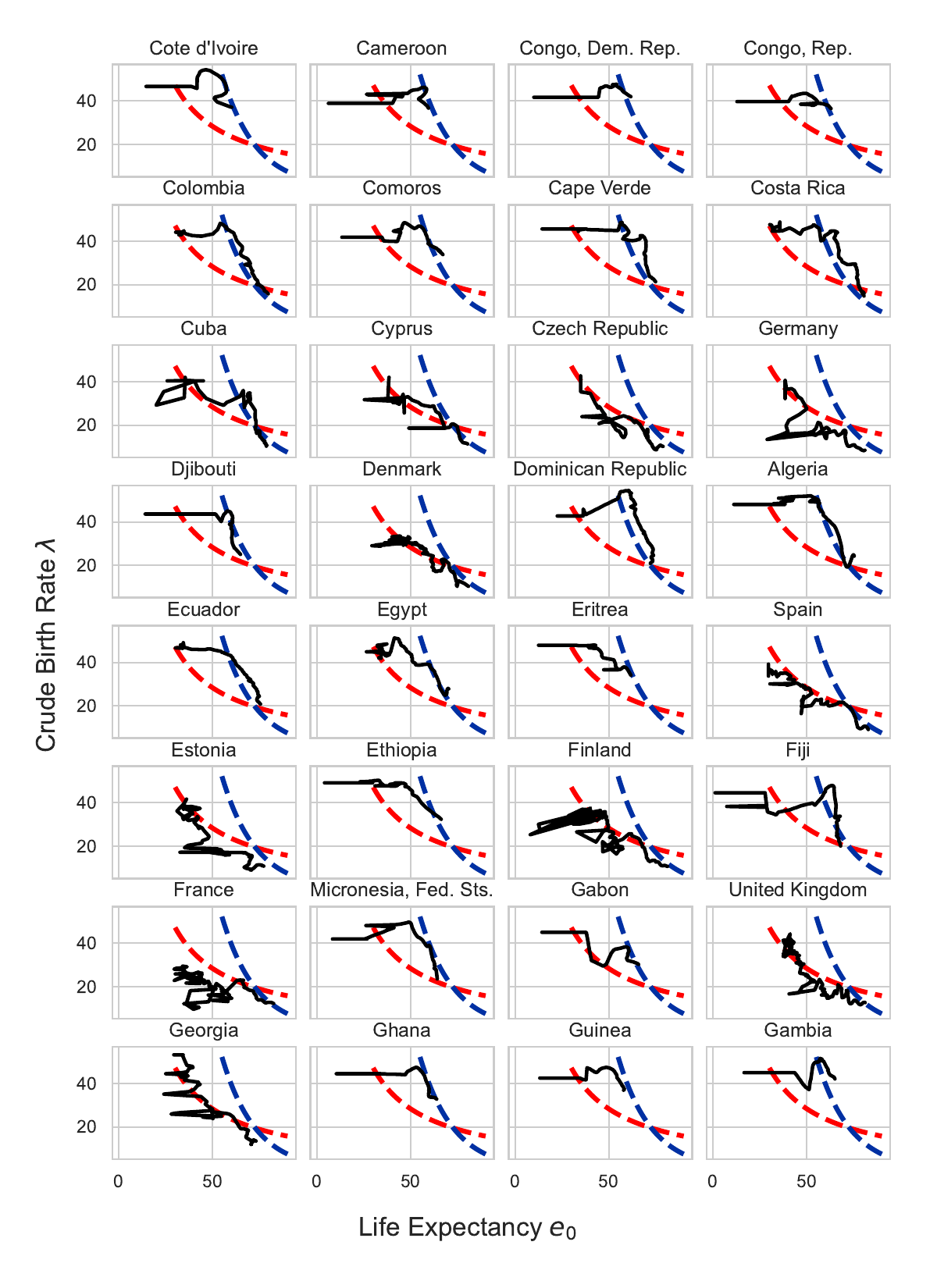}
\caption{Pathways of countries, with two universal pathways as dashed lines.
}
\label{fig:pathway_2}
\end{figure}

\begin{figure}[H]
\centering
\includegraphics[width=.95\textwidth]{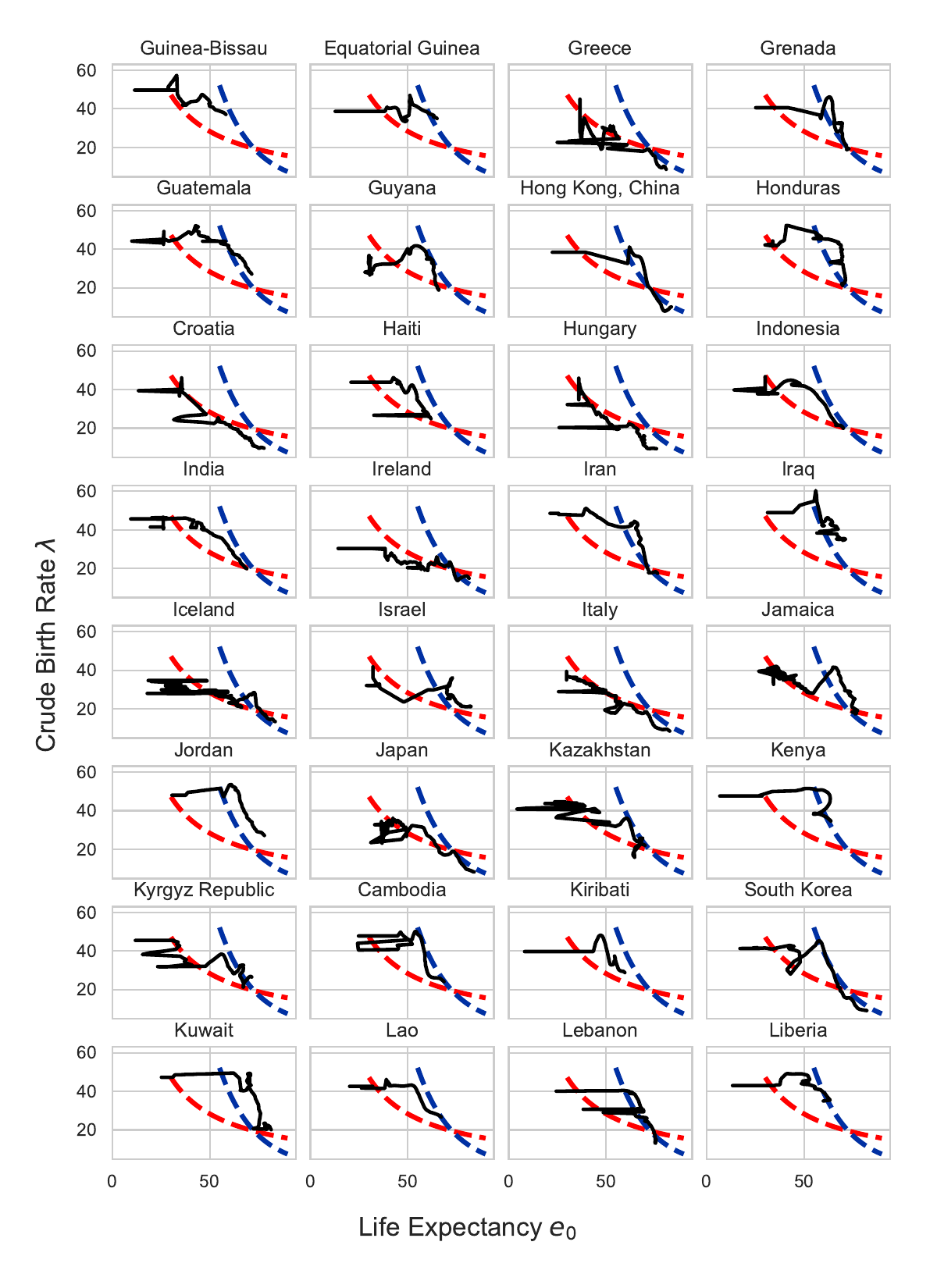}
\caption{Pathways of countries, with two universal pathways as dashed lines.
}
\label{fig:pathway_3}
\end{figure}

\begin{figure}[H]
\centering
\includegraphics[width=.95\textwidth]{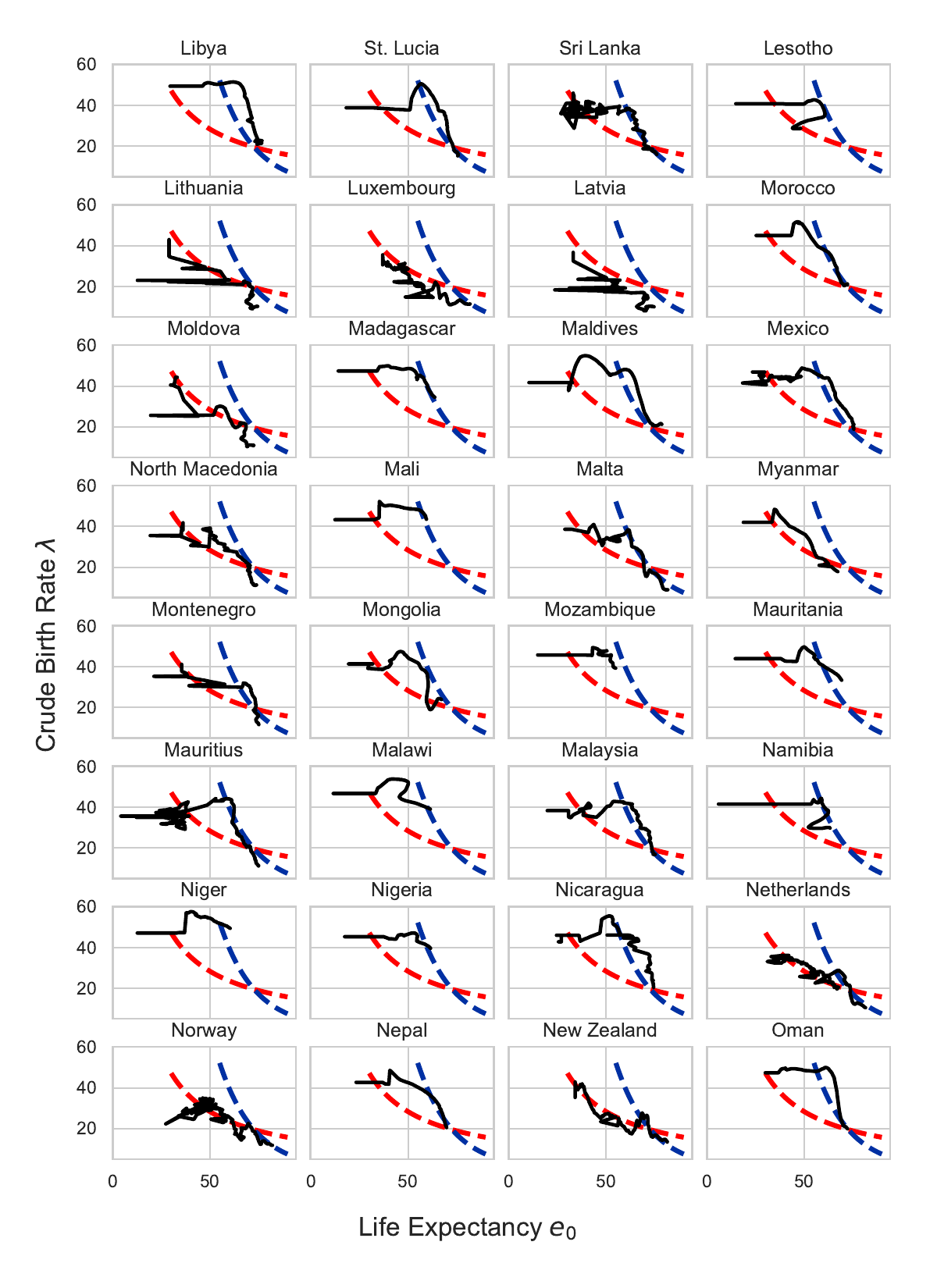}
\caption{Pathways of countries, with two universal pathways as dashed lines.
}
\label{fig:pathway_4}
\end{figure}

\begin{figure}[H]
\centering
\includegraphics[width=.95\textwidth]{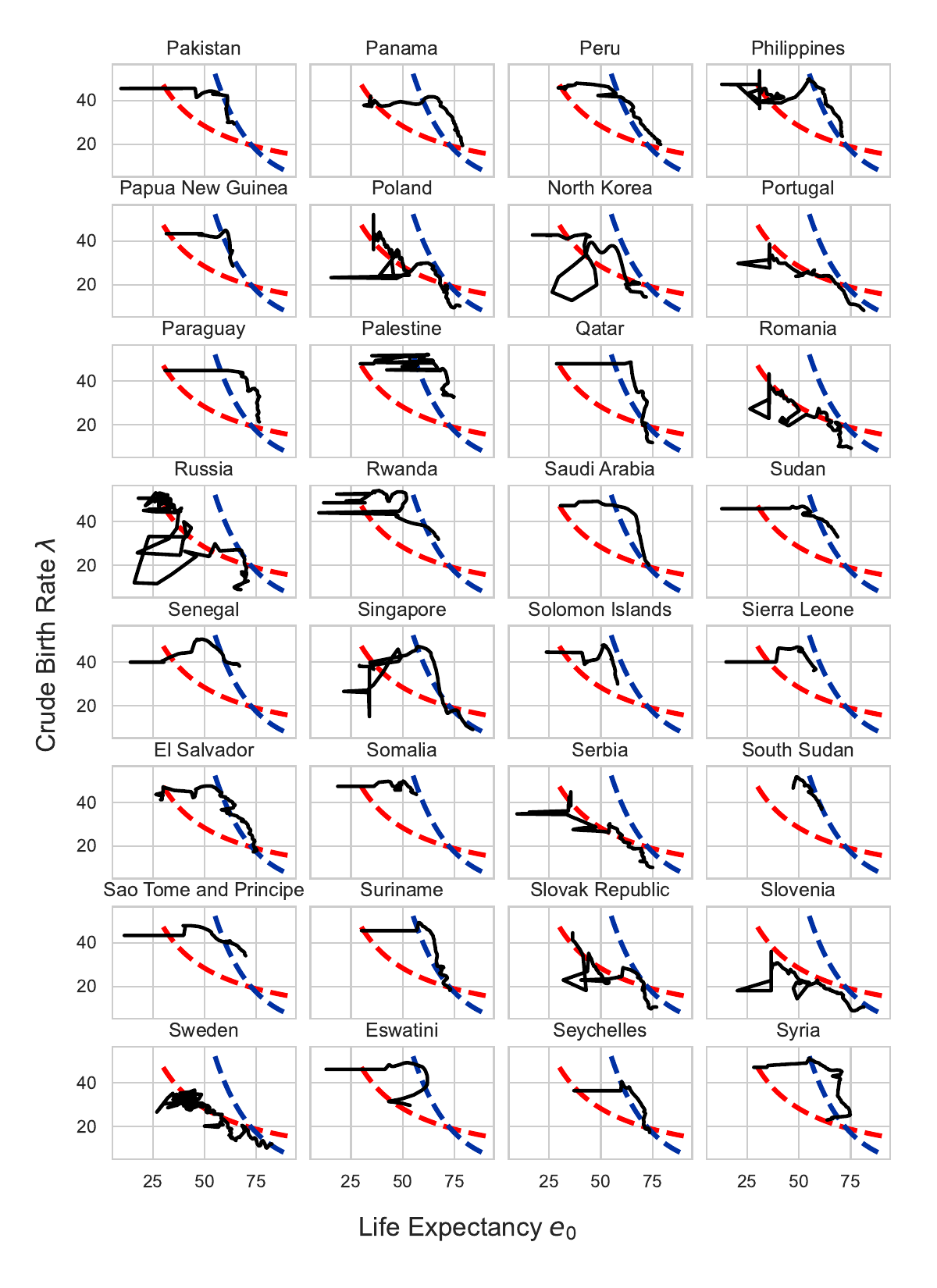}
\caption{Pathways of countries, with two universal pathways as dashed lines.
}
\label{fig:pathway_5}
\end{figure}

\begin{figure}[H]
\centering
\includegraphics[width=.95\textwidth]{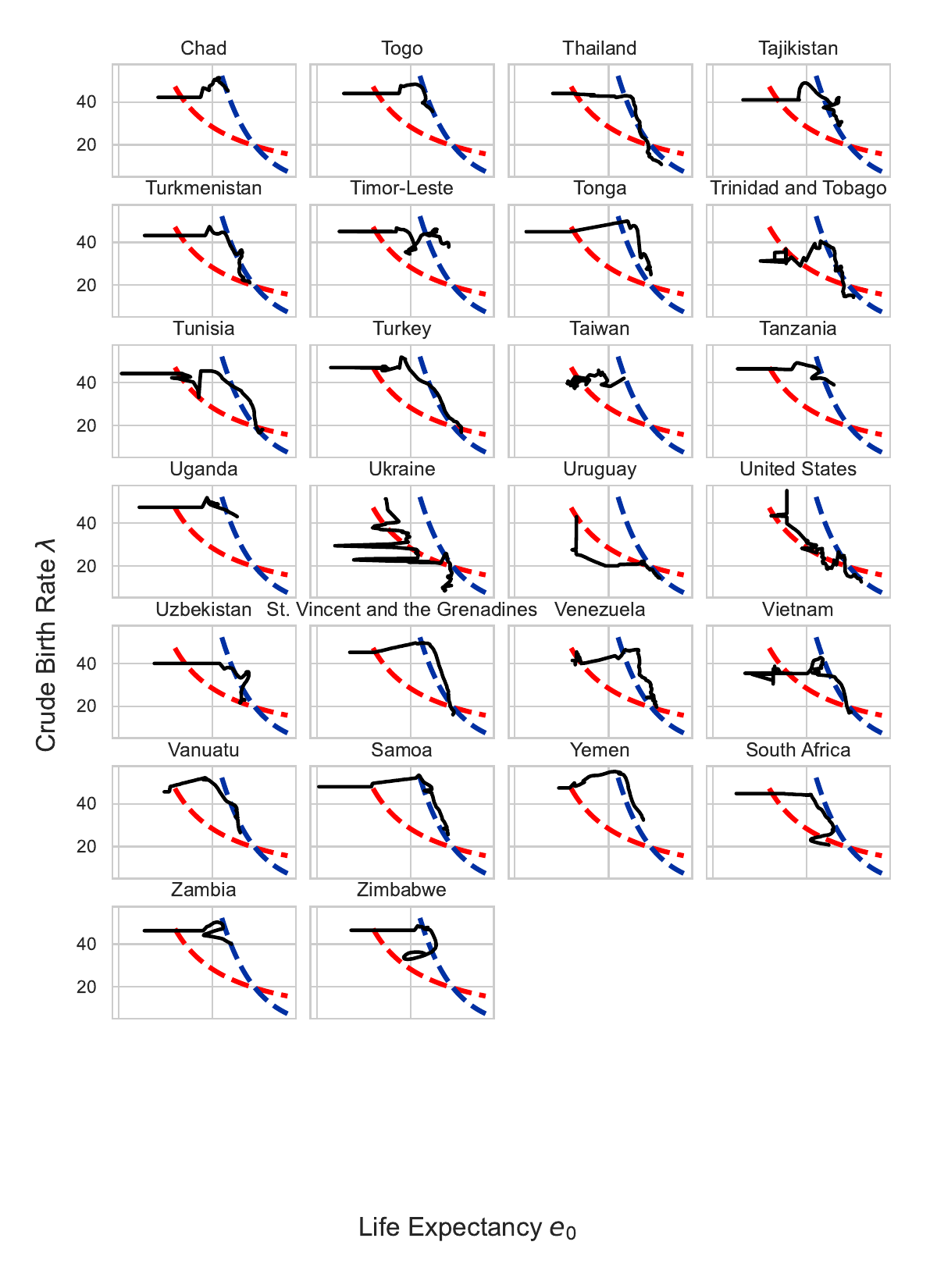}
\caption{Pathways of countries, with two universal pathways as dashed lines.
}
\label{fig:pathway_6}
\end{figure}

\begin{figure}[H]
\centering
\includegraphics[width=.8\textwidth]{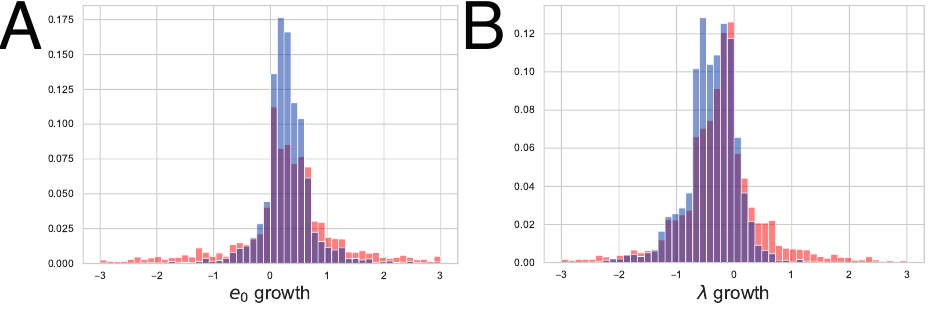}
\caption{
Indices characterizing the two phases of demographic transition. The histograms show the annual increments of (A) life expectancy at birth $e_0$ and (B) the crude birth rate $\lambda$.
}
\label{fig:phase_si}
\end{figure}

\begin{figure}[H]
\centering
\includegraphics[width=\textwidth]{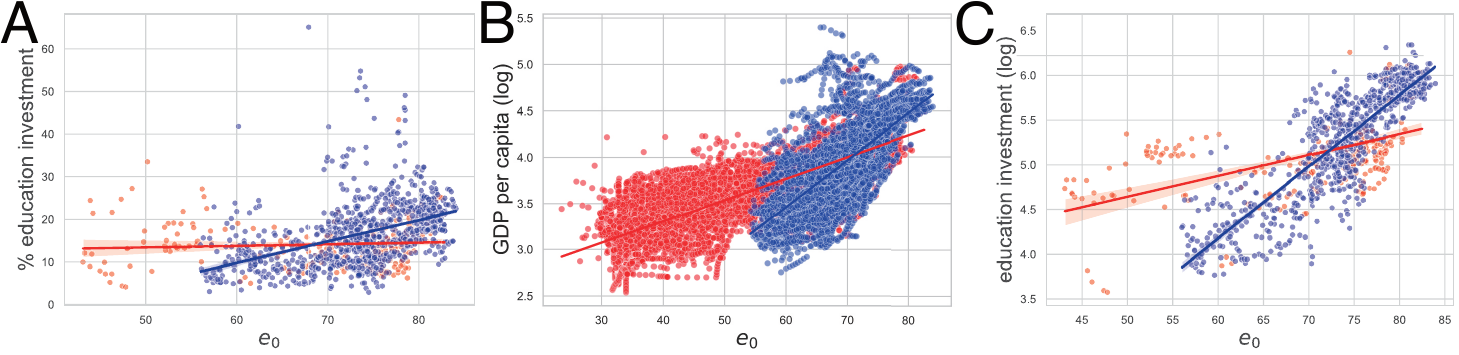}
\caption{Educational investments depending on two phases of demographic transitions. The scatter plot of (A) the percent expenditure on educational investment relative to GDP per capita, (B) GDP per capita, and (C) the education investment per student (log scale, USD), against $e_0$. The Y-axis in (B, C) are log scale. (C) is identical with Fig. 2(F).
Red and blue show the data of phases \one\  and \two, respectively.
Note that many countries shift from Phase \one\ to \two\ at $e_0 = 70$, indicating that educational investment increases much more sharply after this shift. The correlation between the percent expenditure on education and $e_0$ (A) is $0.08$ for Phase \one\ and $0.43$ for Phase \two. That between the logarithm of education investment per student and $e_0$ (C) is $0.56$ for Phase \one\ and $0.86$ for Phase \two.
}
\label{fig:phase_education}
\end{figure}

\begin{figure}[H]
\centering
\includegraphics[width=.5\textwidth]{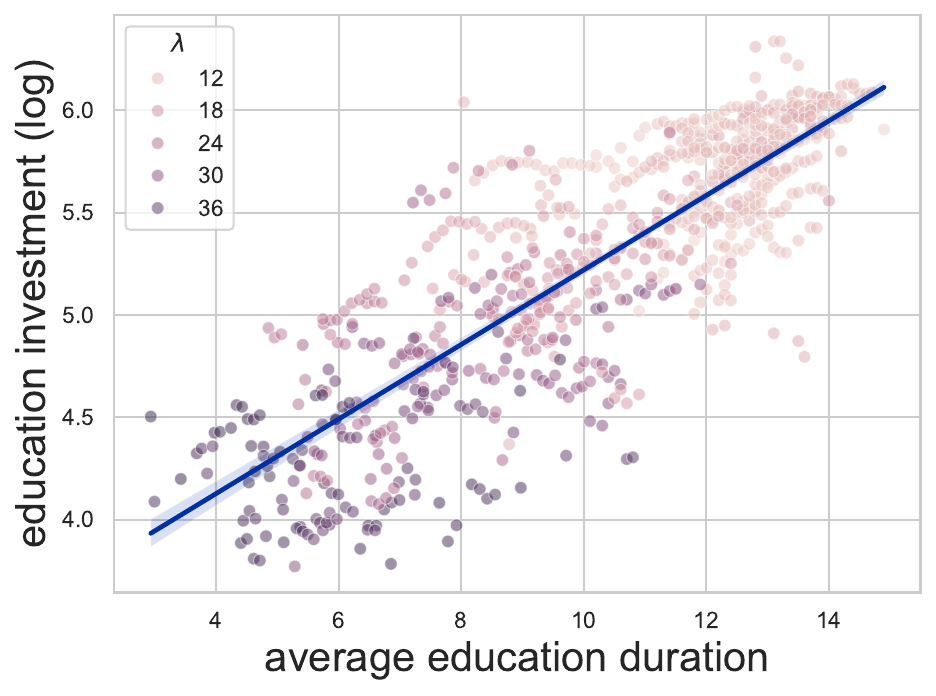}
\caption{
Relationship between average educational duration and education investment per student (log scale, USD). Colors represent the crude birth rate, $\lambda$. The blue line denotes the results of linear regression. Data points below the blue line, representing relatively lower educational costs ($c$), are associated with higher fertility rates, as indicated by the darker colors.
}
\label{fig:lambda_c}
\end{figure}

\newpage
\section*{Supplementary Table}
 \begin{longtable}{p{4.6cm}lp{4.6cm}}
  \caption{The year in which each country experienced each phase of demographic transition. The data for the year until 2015 were used for the analysis.} \label{tab:DT_phase}\\
Phase I	&	Country	&	Phase I\hspace{-1.2pt}I	\\\hline
  \endfirsthead
Phase I	&	Country	&	Phase I\hspace{-1.2pt}I	\\\hline
\endhead
---	&	Afghanistan	&	2009-2015	\\
1936-1941	&	Albania	&	1960-1993, 1997-2015	\\
---	&	Algeria	&	1983-1998, 2004-2014	\\
---	&	Andorra	&	---	\\
---	&	Angola	&	---	\\
1929-1948, 1985-2006	&	Antigua and Barbuda	&	1949-1976, 2007-2015	\\
1921-1951, 1955-1971, 1996-2015	&	Argentina	&	1972-1995	\\
1946-1954, 1968-1975	&	Armenia	&	1957-1967, 1976-1987, 1996-2015	\\
1882-1945, 1966-1973	&	Australia	&	1946-1965, 1974-2015	\\
1897-1943, 1946-1970	&	Austria	&	1971-2015	\\
1993-2015	&	Azerbaijan	&	1965-1992	\\
1931-1949	&	Bahamas	&	1950-1997, 2001-2015	\\
---	&	Bahrain	&	1968-2015	\\
---	&	Bangladesh	&	1988-2013	\\
1929-1947, 1971-1978	&	Barbados	&	1948-1970, 1979-2015	\\
1927-1940, 1946-1951	&	Belarus	&	1952-2015	\\
1807-1841, 1873-1965	&	Belgium	&	1966-2015	\\
---	&	Belize	&	1979-2015	\\
---	&	Benin	&	2005-2015	\\
---	&	Bhutan	&	1988-2008	\\
---	&	Bolivia	&	1982-2015	\\
1930-1940, 1969-1985	&	Bosnia and Herzegovina	&	1954-1968, 1986-1991, 1996-2015	\\
1993-2015	&	Botswana	&	1985-1992	\\
---	&	Brazil	&	1968-2001, 2004-2015	\\
---	&	Brunei	&	1963-2015	\\
1924-1963	&	Bulgaria	&	1964-2015	\\
---	&	Burkina Faso	&	---	\\
---	&	Burundi	&	---	\\
---	&	Cambodia	&	1992-1999, 2005-2015	\\
---	&	Cameroon	&	2009-2015	\\
1864-1944	&	Canada	&	1945-2015	\\
---	&	Cape Verde	&	1972-1979, 1989-2010	\\
2004-2015	&	Central African Republic	&	---	\\
---	&	Chad	&	---	\\
1941-1949, 1991-1998	&	Chile	&	1961-1990, 1999-2015	\\
1976-1984	&	China	&	1966-1975, 1985-1990, 1996-2015	\\
1999-2014	&	Colombia	&	1968-1998	\\
---	&	Comoros	&	1993-2015	\\
---	&	Congo, Dem. Rep.	&	---	\\
---	&	Congo, Rep.	&	2005-2015	\\
1991-2009	&	Costa Rica	&	1965-1990	\\
---	&	Cote d'Ivoire	&	2008-2015	\\
1929-1940, 1946-1972	&	Croatia	&	1973-2015	\\
1922-1945	&	Cuba	&	1946-2015	\\
1899-1948, 1964-1980, 1986-1992	&	Cyprus	&	1949-1963, 1993-2015	\\
1900-1959, 1969-1980	&	Czech Republic	&	1960-1968, 1981-2015	\\
1802-1807, 1815-1827, 1837-1848, 1859-1953	&	Denmark	&	1954-2015	\\
---	&	Djibouti	&	1980-2015	\\
---	&	Dominica	&	---	\\
---	&	Dominican Republic	&	1971-2015	\\
2010-2015	&	Ecuador	&	1971-2009	\\
---	&	Egypt	&	1979-2015	\\
---	&	El Salvador	&	1975-2004	\\
1972-1979	&	Equatorial Guinea	&	2000-2015	\\
---	&	Eritrea	&	2001-2015	\\
1817-1822, 1870-1916, 1920-1963, 1975-1985	&	Estonia	&	1964-1971, 1986-2015	\\
1998-2015	&	Eswatini	&	1992-1997	\\
---	&	Ethiopia	&	2005-2015	\\
2005-2015	&	Fiji	&	1964-2004	\\
1838-1845, 1882-1968	&	Finland	&	1969-2015	\\
1843-1848, 1860-1869, 1872-1913, 1919-1964	&	France	&	1965-2015	\\
1941-1964	&	Gabon	&	1976-2015	\\
---	&	Gambia	&	---	\\
1934-1940, 1945-1994	&	Georgia	&	1995-2015	\\
1888-1915, 1919-1943, 1946-1967	&	Germany	&	1968-2015	\\
---	&	Ghana	&	1987-2015	\\
1897-1959	&	Greece	&	1960-2015	\\
1999-2004	&	Grenada	&	1946-1952, 1963-1998	\\
---	&	Guatemala	&	1986-2015	\\
---	&	Guinea	&	2009-2015	\\
---	&	Guinea-Bissau	&	---	\\
1933-1943, 2001-2015	&	Guyana	&	1959-2000	\\
2000-2009	&	Haiti	&	1992-1999	\\
---	&	Honduras	&	1984-2013	\\
1926-1941	&	Hong Kong, China	&	1942-1972, 1975-2015	\\
1912-1963, 1967-1981	&	Hungary	&	1982-2015	\\
1884-1942, 1971-1996	&	Iceland	&	1943-1970, 1997-2006, 2010-2015	\\
2001-2015	&	India	&	1980-2000	\\
1992-2015	&	Indonesia	&	1974-1991	\\
---	&	Iran	&	1986-2008	\\
---	&	Iraq	&	1978-2015	\\
1866-1958, 2006-2013	&	Ireland	&	1959-1981, 1984-2005	\\
1937-1944, 1984-2015	&	Israel	&	1945-1983	\\
1896-1915, 1919-1969	&	Italy	&	1970-2015	\\
1931-1949, 1997-2008	&	Jamaica	&	1950-1959, 1964-1996	\\
1910-1944, 1950-1966	&	Japan	&	1967-2015	\\
---	&	Jordan	&	1980-2012	\\
1991-2010	&	Kazakhstan	&	1950-1974, 1979-1990	\\
---	&	Kenya	&	2005-2015	\\
1970-1977, 1996-2015	&	Kiribati	&	---	\\
1992-2015	&	Kuwait	&	1955-1960, 1978-1988	\\
1948-1953, 1999-2006	&	Kyrgyz Republic	&	1961-1998, 2007-2015	\\
---	&	Lao	&	1996-2015	\\
1882-1916, 1919-1940, 1946-1962, 1975-1985	&	Latvia	&	1963-1974, 1986-1992, 1996-2015	\\
---	&	Lebanon	&	1950-1975, 1983-1996, 2001-2015	\\
1997-2015	&	Lesotho	&	1981-1996	\\
---	&	Liberia	&	2005-2015	\\
1997-2010	&	Libya	&	1979-1996	\\
1895-1940, 1945-1959	&	Lithuania	&	1971-1992, 1995-2015	\\
1886-1967	&	Luxembourg	&	1968-2015	\\
---	&	Madagascar	&	2001-2015	\\
---	&	Malawi	&	---	\\
---	&	Malaysia	&	1961-2001, 2004-2015	\\
2003-2015	&	Maldives	&	1991-2002	\\
---	&	Mali	&	---	\\
1909-1929, 1963-1968	&	Malta	&	1930-1939, 1942-1962, 1969-2015	\\
---	&	Marshall Islands	&	---	\\
---	&	Mauritania	&	1990-2015	\\
1981-1997	&	Mauritius	&	1964-1980, 1998-2015	\\
2004-2015	&	Mexico	&	1975-2003	\\
2006-2015	&	Micronesia, Fed. Sts.	&	1973-2005	\\
1948-1997	&	Moldova	&	1998-2015	\\
---	&	Monaco	&	---	\\
1992-2011	&	Mongolia	&	1979-1991	\\
1922-1939	&	Montenegro	&	1947-1971, 1976-2015	\\
---	&	Morocco	&	1976-2000, 2006-2015	\\
---	&	Mozambique	&	---	\\
1981-2015	&	Myanmar	&	---	\\
1999-2007	&	Namibia	&	1984-1998, 2008-2015	\\
---	&	Nauru	&	---	\\
---	&	Nepal	&	1989-2011	\\
1876-1945	&	Netherlands	&	1946-1963, 1968-2015	\\
1884-1940	&	New Zealand	&	1945-2015	\\
---	&	Nicaragua	&	1985-2010	\\
---	&	Niger	&	---	\\
---	&	Nigeria	&	---	\\
1944-1949, 1974-2006	&	North Korea	&	2007-2015	\\
1931-1949, 1959-1964, 1983-1994	&	North Macedonia	&	1965-1980, 1995-2015	\\
1814-1945	&	Norway	&	1952-2015	\\
---	&	Oman	&	1989-2014	\\
---	&	Pakistan	&	1990-2015	\\
---	&	Palau	&	---	\\
---	&	Palestine	&	1997-2015	\\
1991-2015	&	Panama	&	1962-1990	\\
---	&	Papua New Guinea	&	1979-2015	\\
2005-2015	&	Paraguay	&	1961-2004	\\
2001-2015	&	Peru	&	1973-2000	\\
---	&	Philippines	&	1967-2015	\\
1903-1941, 1946-1951, 1961-1987	&	Poland	&	1952-1960, 1988-2015	\\
1929-1978	&	Portugal	&	1979-2015	\\
1996-2002	&	Qatar	&	1963-1992, 2003-2015	\\
1929-1989	&	Romania	&	1990-2015	\\
1946-1951, 1962-1989, 1999-2008	&	Russia	&	1952-1961, 2009-2015	\\
---	&	Rwanda	&	2005-2015	\\
---	&	Samoa	&	1971-2015	\\
---	&	San Marino	&	---	\\
---	&	Sao Tome and Principe	&	1972-2015	\\
---	&	Saudi Arabia	&	1984-2014	\\
---	&	Senegal	&	1996-2015	\\
1929-1941, 1945-1989	&	Serbia	&	1990-2015	\\
1881-1908, 1995-2006	&	Seychelles	&	1933-1994, 2007-2015	\\
---	&	Sierra Leone	&	---	\\
---	&	Singapore	&	1959-1972, 1978-1989, 1994-2015	\\
1908-1949, 1962-1985	&	Slovak Republic	&	1950-1961, 1986-2015	\\
1924-1979	&	Slovenia	&	1980-2015	\\
---	&	Solomon Islands	&	1994-2012	\\
---	&	Somalia	&	---	\\
1995-2015	&	South Africa	&	1969-1994	\\
1949-1952	&	South Korea	&	1963-2015	\\
---	&	South Sudan	&	2005-2015	\\
1912-1917, 1921-1972	&	Spain	&	1973-2015	\\
1989-2000	&	Sri Lanka	&	1950-1988, 2005-2015	\\
---	&	St. Kitts and Nevis	&	---	\\
---	&	St. Lucia	&	1968-2015	\\
---	&	St. Vincent and the Grenadines	&	1970-1999, 2004-2015	\\
---	&	Sudan	&	1996-2015	\\
2002-2010	&	Suriname	&	1967-2001	\\
1822-1827, 1840-1852, 1858-1949	&	Sweden	&	1950-2015	\\
1878-1963	&	Switzerland	&	1964-2015	\\
---	&	Syria	&	1984-2012	\\
---	&	Taiwan	&	---	\\
---	&	Tajikistan	&	1972-1983, 1989-2015	\\
---	&	Tanzania	&	---	\\
1988-1996	&	Thailand	&	1967-1987, 1997-2015	\\
---	&	Timor-Leste	&	2002-2015	\\
---	&	Togo	&	2006-2015	\\
---	&	Tonga	&	1967-2015	\\
1923-1940, 1990-1999	&	Trinidad and Tobago	&	1952-1989, 2000-2015	\\
2010-2015	&	Tunisia	&	1972-2009	\\
2000-2013	&	Turkey	&	1971-1999	\\
1999-2009	&	Turkmenistan	&	1966-1998	\\
---	&	Tuvalu	&	---	\\
---	&	Uganda	&	---	\\
1923-1931, 1934-1940	&	Ukraine	&	1950-1960, 1964-1977, 1985-2015	\\
1994-2005	&	United Arab Emirates	&	1968-1993, 2006-2015	\\
1827-1839, 1865-1965	&	United Kingdom	&	1966-2015	\\
1882-1945, 1964-1970	&	United States	&	1946-1963, 1971-2015	\\
1931-1960, 1963-1984	&	Uruguay	&	1985-2015	\\
1998-2015	&	Uzbekistan	&	1950-1997	\\
---	&	Vanuatu	&	1974-2015	\\
2004-2015	&	Venezuela	&	1968-2003	\\
---	&	Vietnam	&	1973-1996, 2000-2015	\\
---	&	Yemen	&	1999-2015	\\
---	&	Zambia	&	---	\\
1995-2008	&	Zimbabwe	&	1987-1994	
 \end{longtable}

\end{document}